\documentstyle[aps,manuscript]{revtex}
\tightenlines
\draft
\begin{document}
\title{A scalar field governed cosmological
model from noncompact Kaluza-Klein theory}
\author{$^1$Jose Edgar Madriz
Aguilar\footnote{E-mail address: edgar@itzel.ifm.umich.mx},
$^{2}$Mauricio Bellini\footnote{E-mail address: mbellini@mdp.edu.ar}
and $^1$Francisco Astorga Saenz\footnote{E-mail address:
astorga@ifm1.ifm.umich.mx}}
\address{$^1$Instituto de F\'{\i}sica y Matem\'aticas, AP: 2-82, (58040)
Universidad Michoacana de San Nicol\'as de Hidalgo, Morelia,
Michoac\'an, M\'exico.\\ $^2$Consejo Nacional de Investigaciones
Cient\'{\i}ficas y T\'ecnicas (CONICET) and Departamento de
F\'{\i}sica, Facultad de Ciencias Exactas y Naturales, Universidad
Nacional de Mar del Plata, Funes 3350, (7600) Mar del Plata,
Argentina.}
\vskip .5cm
\maketitle
\begin{abstract}
This paper is a review of a recently introduced cosmological model from
a noncompact Kaluza-Klein theory for a single scalar field minimally
coupled to gravity. We obtain that the 4D scalar potential has a geometrical
origin and assume different representations in different frames. It should
be responsible for the expansion of the universe. In this framework we explain
the (neutral scalar field governed) evolution of the universe from an
initially inflationary expansion that has a change of phase towards
a decelerated expansion and thereinafter evolves towards the
present day observed accelerated (quintessential) expansion. Finally,
using the Hamilton-Jacobi formalism, we study extra force and extra mass
from this 5D cosmological model.
\end{abstract}
\vskip .2cm
\vskip 2cm
\section{Introduction}

The idea that the Universe may have more than 4 dimensions is due to
Kaluza (1921), who with a brilliant insight realized that a 5D manifold
could be used to unify Einstein's theory of general relativity with
Maxwell's theory of electromagnetism. After some delay, Einstein
endorsed the idea, but a major impetus was provided by Klein (1926). He
made the connection to quantum theory by assuming  that the extra
dimension was microscopically small, with a size in fact
connected via the Planck's constant to the magnitude of the electron charge.
The development of particle physics, quantum field theory and the strings
theory led to a resurgence of interest in higher dimensional field theories as
a means of unifying the long range and short range interactions of
physics. Thus Kaluza-Klein 5D theory lay the foundation for modern
developments such as 10D superstrings and 11D supergravity. There are
several versions of this theory such as noncompactified induced
matter or space-time-matter theory.
The Kaluza-Klein theory is essentially general relativity  in 5D,
and physically have the motivation of explaining why we
perceive 4 dimensions of the space-time and (apparently) do not
see the fifth dimension. It is constrained by two
conditions. (1) The so called ``cylinder condition''
was introduced by Kaluza,
and consists in setting all partial derivatives with respect to
the fifth coordinate to zero. (2) The condition of compactification was
introduced by Klein, and consists in the assumption that the
fifth dimension is not small in size but has a closed
topology (a circle if we are only considering one extra dimension). It
is a constraint that may be applied retroactively to a solution.
This condition introduces periodicity and allows one to use
Fourier and other descompositions of the theory.
The field equations would logically be expected to be $G_{AB}=kT_{AB}$
(where $A,B =0,1,2,3,4$)
with some appropiate coupling constant $k$ and a $5D$ energy momentum tensor.
From the time of Kaluza-Klein onward much
work has been done with the ``apparent vacuum'' or ``empty'' form of the
field equations $G_{AB}=0$. In the practice is very difficult determine that
relations without some starting assumption about $g_{AB}$. This is usually
connected with the physical situation being investigated. In gravitational
theory, an assumption about $g_{AB}=g_{AB}(x^c)$ is commonly called a
choice of coordinates, while in particle physics it is commonly
called a choice of gauge. The traditional Kaluza-Klein theory
has been worked by many people, including Jordan\cite{Jor1,Jor2}, Bergmann\cite{Ber},
Lessner\cite{Less}, Thiry\cite{Thi}, and Liu and Wesson\cite{LW}.
In this theory the coordinates are chosen so as to write
the 5D metric tensor in the form
\begin{equation}\label{eq:p1}
g_{AB}=\left( \begin{array}{cc}
g_{\alpha \beta} -
k^{2}\Phi^{2} A_{\alpha} A_{\beta} & -k \Phi^{2} A_{\alpha} \\
-k \Phi^{2} A_{\beta} & -\Phi^{2} \end{array} \right)  \ ,
\end{equation}
where $k$ is a coupling constant. Then the Einstein's field equations in
the vacuum reduce to:
\begin{equation}
G_{\mu \nu}=\frac{k^{2}\Phi^{2}}{2}T_{\mu\nu}-
\frac{1}{\Phi}(\nabla_{\mu}\nabla_{\nu}\Phi -g_{\mu\nu}\Box\Phi),
\label{eq:p2}
\end{equation}
\begin{equation}
\nabla^{\mu}F_{\mu\nu}=-3\frac{\nabla^{\mu}\Phi}{\Phi}F_{\mu\nu},
\label{eq:p3}
\end{equation}
\begin{equation}
\Box\Phi=-\frac{k^{2}\Phi^3}{4}F_{\mu\nu}F^{\mu\nu},
\label{eq:p4}
\end{equation}
where $\mu, \nu=0,1,2,3$.
In these equations $G_{\mu\nu}$ is the Einstein's tensor, $F_{\mu\nu}$
is the Maxwell's tensor and $T_{\mu\nu}$ is the energy-momentum tensor
for an electromagnetic field  given by $T_{\mu\nu}=
{1\over 2}(g_{\mu\nu}F_{\alpha\beta}{F^{\alpha\beta}\over 4}
-F^{\gamma}\,_{\mu}F_{\nu \gamma})$. Also $\Box\equiv
g^{\mu\nu}\nabla_{\mu}\nabla_{\nu}$ is the wave operator, and
summation convention is in effect. The
equations (\ref{eq:p3}) are the 4 equations of electromagnetism modified
by a function, which by (\ref{eq:p4}) can be thought of as
depending on a wave-like scalar field. The right side of (\ref{eq:p2})
in some sense represents an energy-momentum tensor that is
effectively derived from the fifth dimension. In short the traditional
Kaluza-Klein theory is in general a unified account of gravity,
electromagnetism and scalar field. In the language of particle physics,
the field equations $G_{AB}=0$ of Kaluza-Klein theory describe a spin-2
graviton, a spin-1 photon and a spin-0 boson which is connected with howparticles acquire mass.

\section{Cosmological Model from Induced Matter Theory or Space-Time
Matter theory}

Einstein introduce the idea that the physical quantities should be given
a geometrical interpretation, as envisaged by many people through time.
An early attempt at this was made by Kaluza and Klein, who extended
general relativity from 4 to 5 dimensions, but also applied severe
restrictions to the geometry (the condition of cylindricity and
compactification). In the 90's Paul Wesson and Ponce de Leon showed
that it is possible to interpret most properties of matter as the result
of 5D Riemannian geometry, where however the latter allows dependence
on the fifth coordinate and does not make assumptions about the
topology of the fifth dimension. This theory is called the induced
matter theory.
The induced matter theory has seen most work in 3 areas: the case
of uniform cosmological models, the soliton case and the case of
neutral matter. The first case is easiest to trate because
of the high degree of symmetry involved. The second case is
more complicated, but important because 5D solitons are the analogs of
isolated 4D masses, and the 5D class of soliton solution contains
the unique 4D Schwarzschild solution. The last case can be
treated quite generally, and lays the foundation for many applications
where electromagnetic effects are not involved. We are going to
give a briefly review of the main features considering only the cosmological
case. The other cases go beyond the scope of this work.

In the cosmological context the extra dimension is already known to be
of great importance for cosmology. There is a class of 5D
cosmological models which reduce to the usual four dimensional ones,
on hypersuperfaces defined by setting the value of the extra
coordinate constant. In these models the matter is explained as the
consequence of geometry in five dimensions. The physics of this
follows from a mathematical results. The basic idea of this models
is explained below.
The 5D Einstein's field equations for apparent vacuum  are:
\begin{equation}
G_{AB}=0,
\label{eq:p5}
\end{equation}
where the 5D Einstein's tensor is $G_{AB}=R_{AB}-{1\over 2}Rg_{AB}$,
with $R_{AB}$ the 5D Ricci's tensor and $g_{AB}$ the 5D metric. The
central thesis of induced matter theory is that from equations (\ref{eq:p5})
we obtain the 4D field equations with matter given by:
\begin{equation}
G_{\mu\nu}=8\pi T_{\mu\nu}.
\label{eq:p6}
\end{equation}
In other words, the equations (\ref{eq:p6}) are a subset of (\ref{eq:p5})
with an effective or induced 4D energy-momentum tensor $T_{\mu\nu}$ which
contains the classical properties of matter. This idea can be explained
as a consequence of the Campell's theorem. It says that any analytic
N-dimensional Riemannian manifold  can be locally embedded in an
(N+1)-dimensional Ricci flat Riemannian manifold\cite{RTZ}.
This is of great importance for establishing the
generality of the proposal that 4D field equations with sources can be
locally embedded in 5D field equations without sources. Besides, it
can be used to study lower dimensional gravity $(N<4)$\cite{RRT}.
It can be employed to find new classes of 5D
solutions\cite{LRTR}. Some of the latter have the
remarkable property that they are 5D flat but contain 4D subspaces
that are curved and correspond to known physical situations\cite{W,ACM}.
However, the principle
is clear: curved 4D physics can be embedded in curved or flat 5D geometry.

In this theory an exact solution of (\ref{eq:p5}) is of cosmological type
if resembles that of Friedmann-Robertson-Walker (FRW), and the dynamics is
governed by equations like those of Friedmann. Paul Wesson, Ponce
de Leon and co-workers founded several classes of exact
cosmological solutions of (\ref{eq:p5}) whose metrics are separable
and reduce to the standard 4D FRW ones on the hypersurfaces with the fifth
coordinate constant.

Following the idea suggested by Wesson and co-workers and to
illustrate the transition from 5D field equations (\ref{eq:p5})
for apparent vacuum to the 4D equations (\ref{eq:p6}) with matter,
it is convenient to start considering a 3D spatially,
isotropic and flat spherically symmetric 5D line element:
\begin{equation}
ds^{2}=-e^{\alpha (\psi,t)}dt^{2}+e^{\beta (\psi,t)}dr^{2}+
e^{\gamma (\psi,t)}d\psi^{2},
\label{eq:p7}
\end{equation}
where $dr^{2}=dx^{2}+dy^{2}+dz^{2}$ and $\psi$ is the fifth coordinate. We
assume that $e^{\alpha}$, $e^{\beta}$ and $e^{\gamma}$ are separable
functions of the variables $\psi$ and $t$. The equations for the
relevant Einstein's elements are:
\begin{equation}
G^{0}_{0}=-e^{-\alpha}\left[\frac{3\dot{\beta}^{2}}{4}+
\frac{3\dot{\beta}\dot{\gamma}}{4}\right]+
e^{-\gamma}
\left[\frac{3\stackrel{\star\star}{\beta}}{2}+
\frac{3\stackrel{\star}{\beta}^2}{2}-
\frac{3\stackrel{\star}{\gamma}\stackrel{\star}{\beta}}{4}\right],
\label{eq:p8}
\end{equation}
\begin{equation}
G^{0}_{4}=e^{-\alpha}\left[\frac{3\stackrel{\cdot\star}{\beta}}{2}
+\frac{3\dot{\beta}\stackrel{\star}{\beta}}{4}-
\frac{3\dot{\beta}\stackrel{\star}{\alpha}}{4}-
\frac{3\stackrel{\star}{\gamma}\dot{\gamma}}{4}\right],
\label{eq:p9}
\end{equation}
\begin{eqnarray}
G^{i}_{i}&=&-e^{-\alpha}\left[\ddot{\beta}+
\frac{3\dot{\beta}^2}{4}+\frac{\ddot{\gamma}}{2}+
\frac{\dot{\gamma}^2}{4}+\frac{\dot{\beta}\dot{\gamma}}{2}-
\frac{\dot{\alpha}\dot{\beta}}{2}-\frac{\dot{\alpha}\dot{\gamma}}{4}\right]
 \nonumber \\
&+& e^{-\gamma}\left[\stackrel{\star\star}{\beta}
+\frac{3\stackrel{\star}{\beta}^2}{4}+
\frac{\stackrel{\star\star}{\alpha}}{2}+
\frac{\stackrel{\star}{\alpha}^2}{4}+
\frac{\stackrel{\star}{\beta}\stackrel{\star}{\alpha}}{2}-
\frac{\stackrel{\star}{\gamma}\stackrel{\star}{\beta}}{2}-
\frac{\stackrel{\star}{\alpha}\stackrel{\star}{\gamma}}{4}\right],
\label{eq:p10}
\end{eqnarray}
\begin{equation}
G^{4}_{4}=-e^{-\alpha}\left[\frac{3\ddot{\beta}}{2}+
\frac{3\dot{\beta}^2}{2}-\frac{3\dot{\alpha}\dot{\beta}}{4}\right]+
e^{-\gamma}\left[\frac{3\stackrel{\star}{\beta}^2}{4}+
\frac{3\stackrel{\star}{\beta}\stackrel{\star}{\alpha}}{4}\right],
\label{eq:p11}
\end{equation}
where the overstar and the overdot denote, respectively, $\partial/\partial
\psi$ and $\partial/\partial t$, and $i=1,2,3$. Following the convention
$(-,+,+,+)$ for the 4D metric, we define $T^{0}_{0}=-\rho_{t}$
and $T^{i}_{i}=P$ (we are considering a 3D isotropic and homogeneous
universe), where $\rho_t$ is the total energy density
and $P$ is the pressure. The 5D vacuum conditions (\ref{eq:p5})
are given by [17]:
\begin{equation}
8\pi G\rho_{t}=\frac{3}{4}e^{-\alpha}\dot{\beta}^2,
\label{eq:12}
\end{equation}
\begin{equation}
8\pi G P=e^{-\alpha}\left[\frac{\dot{\alpha}\dot{\beta}}{2}
-\ddot{\beta}-\frac{3\dot{\beta}^2}{4}\right],
\label{eq:13}
\end{equation}
\begin{equation}
e^{\alpha}\left[\frac{3\stackrel{\star}{\beta}^2}{4}
+\frac{3\stackrel{\star}{\beta}\stackrel{\star}{\alpha}}{4}\right]=
e^{\gamma}\left[\frac{\ddot{\beta}}{2}+\frac{3\dot{\beta}^2}{2}-
\frac{\dot{\alpha}\dot{\beta}}{4}\right],
\label{eq:14}
\end{equation}
where $G$ is the gravitational constant.
Hence, from the equations (\ref{eq:12}) and (\ref{eq:13}) and
taking $\dot{\alpha}=0$, we obtain the equation of state for
the induced matter:
\begin{equation}
P=-\left(\frac{4}{3}\frac{\ddot{\beta}}{\dot{\beta}^2}+1\right)\rho_t.
\label{eq:15}
\end{equation}
Notice that for $\ddot{\beta}/\dot{\beta}^2\le 0$ and
$\mid\ddot{\beta}/\dot{\beta}^2\mid\ll 1$ (or zero), this
equation describes an inflationary universe. The equality
$\ddot{\beta}/\dot{\beta}^2=0$ corresponds with a 4D de Sitter
expansion for the universe. This theory is gauge depending because
for different choice of coordinates, one have several metrics all of
them solutions of (\ref{eq:p5}). In 1988 Ponce de Leon obtained
one of the classes of solutions to (\ref{eq:p5}) which are
solutions of cosmological and astrophysical importance.  With those
line elements is possible to develop models that reduce to the
standard FRW ones with flat 3D space sections on hypersurfaces
$\psi=const.$\cite{librowesson}. This is
one of the most interesting aspects of this theory because one
can ensure that the 5D models reduce to the 4D on
hypersurfaces $x^{4}=const.$. Taking the solution
$e^{\alpha}=\psi^2$, $e^{\beta}=t^{\frac{2}{\alpha}}
\psi^{\frac{2}{1-2\alpha}}$, $e^{\gamma}=\alpha^{2}(1-\alpha)^{-2}t^{2}$,
the basic line element (\ref{eq:p7}) can be written:
\begin{equation}
ds^{2}=-\psi^{2}dt^{2}+t^{2/\alpha}\psi^{2/(1-\alpha)}
[dr^{2}+r^{2}(d\theta^{2}+sin^{2}\theta d\theta^{2}]+
\alpha^{2}(1-\alpha)^{-2}t^{2}d\psi^{2},
\label{eq:p16}
\end{equation}
where $\alpha$ is a constant related in the Space-Time-Matter (STM)
theory
to the properties of matter. This constant is determined by induced energy
momentum tensor related at the theory. From the Einstein's equations
and eq. (\ref{eq:p16}) the equations of state are:
\begin{equation}
8\pi\rho_t =\frac{3}{\alpha^{2}\psi^{2}t^{2}},\quad 8\pi
P=\frac{(2\alpha-3)}{\alpha^{2}\psi^{2}t^{2}}, \quad
P=\left(\frac{2\alpha}{3}-1\right) \rho_t.
\label{eq:p17}
\end{equation}
The choice $\alpha=2$ gives $P=\frac{\rho_t}{3}$ which is typical
of radiation, and a scale factor $a(t)\sim t^{1/2}$.
The choice $\alpha=3/2$ gives $P=0$ which is typical of dust,
and a scale factor that grows as $t^{2/3}$. Thus on hypersurfaces
$\psi=const.$ the standard models for the early and late universe
are recovered.

The coordinates in (\ref{eq:p16}) are spatially comoving as in the
usual presentation of the 4D models. That is, $u^{i}=\frac{dx^i}{ds}=0$.
The other components can be found by solving the
5D geodesic equation to be:
\begin{equation}
u^{0}=\mp\frac{\alpha}{\sqrt{2\alpha-1}}\frac{1}{\psi},
\quad u^{4}=\pm\frac{(1-\alpha)^2}{\alpha\sqrt{2\alpha-1}}\frac{1}{t}.
\label{eq:p18}
\end{equation}
If we now change coordinates to:
\begin{equation}
T=t\psi,\quad R=t^{1/\alpha},\quad \Psi=At^{\pm A}\psi,
\label{eq:p19}
\end{equation}
we find $u^{2}=u^{3}=u^{4}=0$ and:
\begin{equation}
u^{0}=\mp\frac{\sqrt{2\alpha-1}}{\alpha},
\quad u^{1}=\mp\frac{1}{\sqrt{2\alpha-1}}\frac{R}{T},
\label{eq:p20}
\end{equation}
where $g^{00}=(2\alpha-1)/\alpha^2={\rm const}$. The density and
pressure (\ref{eq:p17}) change to:
\begin{equation}
8\pi\rho_t=\frac{3}{\alpha^{2}T^{2}},\quad 8\pi
P=\frac{2\alpha-3}{\alpha^{2}T^{2}}.
\label{eq:p21}
\end{equation}
Energy density and pressure are identical to their 4D values for
radiation and dust, without $\psi$ factor. The presence or absence
of the latter, and the question of whether $u^{4}$ is zero or not,
clearly depends on the choice of coordinates. So the functional
form of $\rho_t$ and $P$ can change depending on the choice
of the fifth coordinate. This feature means that a 5D model
may take different 4D guises depending on the coordinate frame.
A particularly interesting consequence of 5D covariance may be derived
by considering the simple coordinate transformation:
\begin{eqnarray}
T&=&\left(\frac{\alpha}{2}\right)t^{1/\alpha}\psi^{1/(1-\alpha)}
\left(1+\frac{r^2}{\alpha^2}\right)-\frac{\alpha}{2(1-2\alpha)}
\left[t^{-1}\psi^{\alpha/(1-\alpha)}\right]^{(1-2\alpha)/\alpha}, \nonumber \\
R&=&rt^{1/\alpha}\psi^{1/(1-\alpha)},\nonumber \\
\Psi&=&\left(\frac{\alpha}{2}\right)t^{1/\alpha}\psi^{1/(1-\alpha)}
\left(1-\frac{r^2}{\alpha^2}\right)+\frac{\alpha}{2(1-2\alpha)}
\left[t^{-1}\psi^{\alpha/(1-\alpha)}\right]^{(1-2\alpha)/\alpha}.
\label{eq:p22}
\end{eqnarray}
Then the line element (\ref{eq:p16}) becomes:
\begin{equation}
ds^{2}=-dT^{2}+dR^{2}+R^{2}(d\theta^{2}+sin^{2}\theta d\phi^{2})-d\Psi^{2},
\label{eq:p23}
\end{equation}
which is manifestly flat. This surprising result may be verified by computer,
which shows that all of the components of the Riemann-Christoffel tensor for
the 5D metric (\ref{eq:p16}) are zero. Despite this, the model's 4D
part is not flat, since the 4D Ricci scalar may be calculated to
be $6(\alpha-2)/(\alpha^{2}t^{2}\psi^{2})$. We see that while the
universe may be curved in 4D, it is flat in 5D. If we have a
conclusion in the space time matter theory is that one extra
dimension is enough to explain the phenomenological properties of classical
matter. For a complete treatment you must to see\cite{librowesson}.

\section{The evolution of the Universe from non\-com\-pact kaluza klein
theory.}

In a cosmological context, the energy density of scalar fields
has been reconized to contribute to the expansion of the
universe\cite{jd}, and has been proposed to explain inflation\cite{Guth},
as well as the presently observed accelerated expansion\cite{we}.
The observed isotropy and homogeneity of the universe do not allow
for the existence of long-range electric and magnetic fields, but neutral
scalar fields can have non-trivial dynamics in an expanding FRW-type
universe. An attempt to confront the data with the predictions for
a minimally coupled scalar field with an a priori unknown potential was made
recently\cite{sta}.

A very important question in theoretical physics consists to provide
a good geometrical description of matter using only one extra
coordinate (say $\psi$). The explanation of this issue in the
framework of the early
universe, in particular for inflationary theory\cite{Guth}, should be of
great importance in cosmology. In this section, we are aimed
to study this topic using the Kaluza-Klein formalism where
the fifth coordinate is noncompact. In this framework should be
interesting to explain the origin of an effective four dimensional (4D)
scalar potential $V(\varphi)$ which could be originated from a 5D
apparent vacuum.
For example, an attempt to understand inflation [which
is governed by the neutral scalar (inflaton) field], from a 5D
flat Riemannian manifold was made in\cite{NPB}. During inflation,
the scale factor of the universe accelerates and this acceleration is
driven by the potential energy related with the self-interactions
of a scalar field. However, Campell's theorem implies
that all inflationary solutions can be generated, at least in
principle, from a 5D vacuum Einstein gravity\cite{librowesson}.
But, could be possible
to develop a formalism to describe all the evolution of the universe?
The other aim of this section
consists to develop a 5D mechanism inspired in
the Campbell's theorem, to explain the (neutral scalar field governed)
evolution of the universe from a initially inflationary (superluminical)
expansion that has a change of phase towards a decelerated (radiation
and later matter dominated) expansion and thereinafter evolves
towards the present day observed accelerated expansion
(quintessence)\cite{ps}. Although Campell's theorem
relates N-dimensional theories to vacuum $(N+1)$-dimensional
theories, it does not establish a strict equivalence between them. 
It is therefore important to determinate when such teories
are equivalent. Two notions of equivalence that could be considered
are dynamical equivalence and geodesic equivalence. Dynamical equivalence
would imply that the dynamics of vacuum N-dimensional theories is
included in a vacuum $(N+1)$-dimensional theories. Alternatively, one
may consider geodesic equivalence, in the sense on Mashhoon et al.\cite{MLW}.
In this case the $(3+1)$ geodesic equation induces a $(2+1)$
geodesic equation plus a force (per unity of mass) term $F^{C}$:
\[
\frac{dU^{C}}{dS}+\Gamma^{C}\,_{AB}U^{A}U^{B}=F^{C}.
\]
For the geodesic equivalence approach one would therefore require
$F^{C}=0$, that describes the trajectories of free-falling observers.
In this work we shall use the geodesic equivalence approach.

In the last years extra force and extra mass has been subject of study
\cite{14}.
It should be an observable effect from extra dimensions on the
4D spacetime. The final aim of this section is to extend the Hamilton-Jacobi
formalism developed by Ponce de Leon\cite{15}
to cosmological models where
the expansion of the universe is governed by a single inflaton field.
This interpretation has the adventage of being free of the complications
and ambiguities of the geodesic approach.

\subsection{Formalism}

To make it, we consider the 5D metric introduced by Ledesma and
Bellini\cite{PLB}
\begin{equation}\label{6}
dS^2 = \epsilon\left(\psi^2 dN^2 - \psi^2 e^{2N} dr^2 - d\psi^2\right).
\end{equation}
Where $dr^{2}=dx^{2}+dy^{2}+dz^{2}$. Here, the
coordenates ($N$,$r$) are dimensionless, the fifth coordinate $\psi $ has
spatial unities and $\epsilon$ is a dimensionless parameter that can take
the values $\epsilon = 1,-1$. The metric (\ref{6}) describes a flat
5D manifold in apparent vacuum ($G_{AB}=0$). We consider a diagonal
metric because we are dealing only with gravitational effects, which are
the important ones in the global evolution for the universe. To
describe neutral matter in a 5D geometrical vacuum (\ref{6})
we can consider the Lagrangian
\begin{equation}\label{1}
^{(5)}{\rm L}(\varphi,\varphi_{,A}) =
-\sqrt{\left|\frac{^{(5)}
g}{^{(5)}g_0}\right|} \  ^{(5)}{\cal L}(\varphi,\varphi_{,A}),
\end{equation}
where $|^{(5)}g|=\psi^8 e^{6N}$
is the absolute value of the determinant for the 5D metric tensor with
components $g_{AB}$ and
$|^{(5)}g_0|=\psi^8_0 e^{6N_0}$ is a constant of
dimensionalization determined by $|^{(5)}g|$ evaluated at $\psi=\psi_0$
and $N=N_0$.
In this work we shall consider $N_0=0$, so that $^{(5)}g_0=\psi^8_0$.
Here, the index $" 0 "$ denotes the values at the end
of inflation. Furthermore, we shall consider an action
\begin{equation}
I=-\int d^{4}x \  d\psi \  \sqrt{\left|\frac{^{(5)}
g}{^{(5)}g_0}\right|} \ \left[\frac{^{(5)}R}{16\pi G}
+ ^{(5)}{\cal L}(\varphi,\varphi_{,A})\right],
\label{action}
\end{equation}
for a scalar field $\varphi$, which is minimally coupled to gravity. Here,
$^{(5)}R$ es the 5D Ricci scalar, which of course, is zero
for the 5D flat metric (\ref{6}) and $G$ is the gravitational constant.
Since the 5D metric (\ref{6}) describes a manifold in apparent
vacuum, the density Lagrangian ${\cal L}$ in (\ref{1}) is
\begin{equation}\label{1'}
^{(5)}{\cal L}(\varphi,\varphi_{,A}) = 
\frac{1}{2} g^{AB} \varphi_{,A} \varphi_{,B},
\end{equation}
which represents a free scalar field. In other words, we
define the vacuum
as a purely kinetic 5D-lagrangian on a globally 5D-flat metric [in our
case, the metric (\ref{6})]. Taking into account the metric (\ref{6})
and the Lagrangian (\ref{1}), we obtain the equation of
motion for $\varphi$
\begin{equation}\label{df}
\left(2\psi \frac{\partial\psi}{\partial N}+ 3 \psi^2 \right)
\frac{\partial\varphi}{\partial N}
+\psi^2 \frac{\partial^2\varphi}{\partial N^2} - \psi^2
e^{-2N} \nabla^2_r\varphi
-4\psi^3 \frac{\partial\varphi}{\partial\psi} - 3\psi^4 \frac{\partial N}{
\partial\psi} \frac{\partial\varphi}{\partial\psi} - \psi^4 \frac{\partial^2
\varphi}{\partial\psi^2} =0,
\end{equation}
where ${\partial N \over \partial\psi}$ is zero because the coordinates
$(N,\vec{r},\psi)$ are independents.

In this work we shall consider the case where $N=N(t)$.
The relevant Christoffel symbols for the geodesic of the 5D
metric (\ref{6}) in a 3D comoving frame $U^r=0$, are
\begin{equation}
\Gamma^N_{\psi\psi}= 0, \quad \Gamma^N_{\psi N}= 1/\psi, \quad
\Gamma^{\psi}_{NN}= \psi, \quad \Gamma^{\psi}_{N \psi}= 0,
\end{equation}
so that the geodesic dynamics ${dU^C \over dS} = \Gamma^C_{AB} U^A U^B$
is described by the following equations of motion
for the velocities $U^A$
\begin{eqnarray}
&& \frac{dU^{N}}{dS} = -\frac{2}{\psi} U^N U^{\psi}, \\
&& \frac{dU^{\psi}}{dS} = -\psi U^N U^N, \\
&&  \psi^2 U^N U^N - U^{\psi} U^{\psi} =1, \label{geo}
\end{eqnarray}
where the eq. (\ref{geo}) describes the condition
$g_{AB}U^A U^B=1$.
From the general solution $\psi U^N  = {\rm cosh}[S(N)]$,
$U^{\psi}=-{\rm sinh}[S(N)]$, we obtain the equation
that describes the geodesic evolution for $\psi$
\begin{equation}
\frac{d\psi}{dN} = \frac{U^{\psi}}{U^N} = -\psi \  {\rm tanh}[S(N)].
\end{equation}
If we take ${\rm tanh}[S(N)]=-1/u(N)$, we obtain the velocities
$U^A$: 
\begin{equation}\label{tri}
U^{\psi} = - {1 \over \sqrt{u^2(N)-1}}, \qquad U^{r}=0,\qquad
U^N={u(N) \over \psi\sqrt{u^2(N)-1}},
\end{equation}
which are satisfied for $S(N)=\pm |N|$. In this work we shall
consider the case $S(N) = |N|$. In
this representation ${d\psi \over dN}=\psi/u(N)$.
Thus, the fifth coordinate evolves as
\begin{equation}\label{psi}
\psi(N) = \psi_0 e^{\int dN/u(N)}.
\end{equation}
From the mathematical point of view, we are taking a foliation
of the 5D metric (\ref{6}) with $r$ constant. Hence, to describe
the metric in physical coordinates we must to make the following
transformations:
\begin{equation}\label{*}
t = \int \psi(N) dN, \qquad R=r\psi, \qquad L= \psi(N) \  e^{-\int dN/u(N)},
\end{equation}
such that for $\psi(t)=1/h(t)$, we obtain the 5D metric
\begin{equation}
dS^2 = \epsilon\left(dt^2 - e^{2\int h(t) dt} dR^2 - dL^2\right),
\label{nme}
\end{equation}
where $L=\psi_0$ is a constant and $h(t)=\dot b/b$ is the effective
Hubble parameter defined from the effective scale factor of the
universe $b$. The metric (\ref{nme}) describes a 5D generalized FRW metric,
which is 3D spatially flat [i.e., it is flat in terms of $\vec{R}=(X,Y,Z)$],
isotropic and homogeneous. In the representation $(\vec{R},t,L)$,
the velocities $ \hat U^A ={\partial \hat x^A \over \partial x^B} U^B$,
are
\begin{equation} \label{10}
U^t=\frac{2u(t)}{\sqrt{u^2(t)-1}}, \qquad
U^R=-\frac{2r}{\sqrt{u^2(t)-1}}, \qquad U^L=0,
\end{equation}
where the old velocities $U^B$ are $U^N$, $U^r=0$ and $U^{\psi}$
and the velocities $\hat U^B$ are constrained by the condition
\begin{equation}\label{con}
\hat g_{AB} \hat U^A \hat U^B =1.
\end{equation}
Furthermore, the function $u$ can be written as a function of time 
$u(t) = -{h^2 \over \dot h}$,
where the overdot represents the derivative with respect to the time.
The solution $N={\rm arctanh}[1/u(t)]$ corresponds to a
time dependent power-law expanding universe
$h(t)=p_1(t) t^{-1}$, such that the effective scale factor go as
$b \sim e^{\int p_1(t)/t dt}$. When $u^{2}(t)>1$, the velocities $U^t$
and $U^R$ are real, so the condition (\ref{con}) implies that
$\epsilon=1$. [Note that the function $u(t)$ can be related to
the deceleration parameter $q(t)=-\ddot{b}b/\dot{b}^2$: $u(t)=1/[1+q(t)]$].
In such that case the expansion of the universe is accelerated
$(\ddot{b}>0)$. However, when $u^{2}<1$ the velocities $U^t$
and $U^R$ are imaginary and the condition (\ref{con}) holds for
$\epsilon=-1$. In this case the expansion of the universe is decelerated
because $\ddot{b}<0$. So, the parameter $\epsilon$ is introduced
in the metric (\ref{nme}) to preserve the hyperbolic condition (\ref{con}).
Moreover, the coordinates $(\vec{R},t,L)$
has physical meaning, because $t$ is the cosmic time and $(\vec{R},L)$
are spatial variables. Since the line element is a function of
time $t$ (i.e., $S\equiv S(t)$), the new coordinate $R$ give us
the physical distance between galaxies separated by cosmological distances:
$R(t)=r(t)/h(t)$.
Note that for $r >1$ ($r <1$), the 3D spatial distance $R(t)$ is defined
on super (sub) Hubble scales. Furthermore $b(t)$ is the effective
scale factor of the universe and describes its effective 3D
euclidean (spatial) volume. Hence, the effective 4D metric
is a spatially (3D) flat FRW one
\begin{equation}\label{frw}
dS^2 \rightarrow ds^2 = \epsilon
\left(dt^2 - e^{2\int h(t) dt} dR^2\right),
\end{equation}
and has a effective 4D scalar curvature $^{(4)}{\cal R} = 6(\dot h + 2 h^2)$.
The metric (\ref{frw}) has a metric tensor with components $g_{\mu\nu}$.
The absolute value of
the determinant for this tensor is $|^{(4)}g|=(b/b_0)^6$.
Now we can make the same treatment to the density Lagrangian
(\ref{1'}) and the differential equation (\ref{df}). Using the
transformations (\ref{*}) we obtain
\begin{equation}\label{aa}
^{(4)} {\cal L}\left[\varphi(\vec{R},t), \varphi_{,\mu}(\vec{R},t)\right]
= \frac{1}{2} g^{\mu\nu} \varphi_{,\mu} \varphi_{,\nu}
-\frac{1}{2} \left[(R h)^2  -
\frac{b^2_0}{b^2} \right] \  \left(\nabla_R \varphi\right)^2, 
\end{equation}
and the equation of motion for $\varphi$ yields
\begin{equation}\label{bb}
\ddot\varphi  +  3 h\dot\varphi -\frac{b^2_0}{b^2} \nabla^2_R \varphi
+ \left[\left(4\frac{h^3}{\dot h} - 3\frac{\dot h}{h}
-3\frac{h^5}{\dot h^2}\right) \dot\varphi +
\left( \frac{b^2_0}{b^2} - h^2 R^2\right)\nabla^2_R\varphi\right]=0.
\end{equation}
From eqs. (\ref{aa}) and (\ref{bb}), we obtain respectively
the effective scalar 4D potential
$V(\varphi)$ and its derivative with respect
to $\varphi(\vec{R},t)$ are
\begin{eqnarray}
V(\varphi) & \equiv & \frac{1}{2}\left[ (R h)^2
- \left(\frac{b_0}{b}\right)^2 \right] \left(\nabla_R\varphi\right)^2,
\label{au} \\
V'(\varphi) & \equiv  &
 \left(4\frac{h^3}{\dot h} - 3\frac{\dot h}{h} -
3\frac{h^5}{\dot h^2}\right) \dot\varphi 
 + 
\left(\frac{b^2_0}{b^2} - h^2 R^2\right)\nabla^2_R\varphi,\label{a1}
\end{eqnarray}
where the prime denotes the derivative with respect to $\varphi $.
The equations (\ref{aa}) and (\ref{bb}) describe the dynamics of the
inflaton field
$\varphi(\vec{R},t)$ in a metric (\ref{frw}) with a Lagrangian
\begin{equation}\label{l4}
^{(4)}{\cal L}[\varphi(\vec{R},t),\varphi_{,\mu}(\vec{R},t)] =
-\sqrt{\left|\frac{^{(4)}g}{^{(4)}g_0}\right|}
\left[\frac{1}{2} g^{\mu\nu} \varphi_{,\mu}\varphi_{,\nu}
+V(\varphi)\right],
\end{equation}
where $\left|^{(4)}g_0\right|=1$.

In this frame, the 4D energy 
density $\rho_t$ and the pressure $P$ are\cite{PLB}
\begin{eqnarray}
&& 8 \pi G \rho_t = 3 h^2,\\
&& 8\pi G P = -(3h^2 + 2 \dot h).
\end{eqnarray}
From the condition (\ref{con}) we can differenciate some different
stages of the universe. If $u^2(t)={4 r^2 (b/b_0)^2 -1 \over 3} >1$,
we obtain
that $r$ can take the values $r > 1$ ($r < 1$) for
$b/b_0 < 1$ ($b/b_0 > 1$), respectively. In this case $q < 0$,
so that the expansion is accelerated. On the
other hand if $u^2(t)={4 r^2 (b/b_0)^2 -1 \over 3} <1$, $r$
can take the values $r< 1$ ($r > 1$) for
$b/b_0 > 1$ ($b/b_0 < 1$), respectively. In this stage $q >0$
and the expansion of the universe is decelerated, so that the function
$u(t)$ take the values $0 < u(t) <1$ and the velocities (\ref{10})
become imaginary. Thus, the metric (\ref{frw}) shifts its
signature from $(+,-,-,-)$
to $(-,+,+,+)$. When $u(t) =1$ the deceleration parameter becomes
zero because $\ddot b =0$. At this moment the velocities (\ref{10})
rotates
synhcronically in the complex plane and $r$ take
the values $r=1$ or $r<1$, for $b/b_0=1$ or $b/b_0 >1$, respectively.

On the other hand, $V(\varphi)$ and $V'(\varphi)$ can
be written as a function of the old coordinates $(N,r,\psi)$
in the comoving frame $U^r=0$
\begin{eqnarray}
&& V(\varphi) \equiv \frac{1}{2} \left[r^2 - e^{-2N}\right] \frac{1}{r^2}
\left(\frac{1}{\stackrel{\star}{\psi}} \stackrel{\star}{\varphi}\right)^2,
\label{v1} \\
&& V'(\varphi) \equiv
\left(3 \frac{\stackrel{\star}{\psi}}{\psi^3}
-\frac{4}{\psi\stackrel{\star}{\psi}}
-\frac{3}{\stackrel{\star}{\psi}^2}\right)
\stackrel{\star}{\varphi}+
\left[\left(\frac{a_0}{a}\frac{1}{r}\right)^2-1\right] \frac{
\partial^2\varphi}{\partial \psi^2}.\label{v'}
\end{eqnarray}
Here, the overstar denotes
the derivative with respect to $N$.
Note that $\Delta N$ is the number of e-folds of the universe. To inflation
solves the horizon/flatness problems it is required that $\Delta N\ge 60$
at the end of inflation.

At this point we can introduce the 4D
Hamiltonian ${\cal H}=\pi^0 \dot\varphi-
^{(4)}{\rm L}$, where the 4D Lagrangian is
$^{(4)} {\rm L}(\varphi,\varphi_{,\mu})
= \sqrt{{|^{(4)}g|
\over |^{(4)}g_0| }} \  {^{(4)}{\cal L}}(\varphi,\varphi_{,\mu})$
[see eq. (\ref{l4})]:
\begin{equation}
{\cal H} = \frac{1}{2} \frac{a^3}{a^3_0}
\left[ \dot\varphi^2 + \frac{a^2_0}{a^2} \left(\nabla\varphi\right)^2
+2V(\varphi)\right].
\end{equation}
Hence, we can define the effective 4D energy density operator
$\rho_t$ such that
\begin{equation}
\rho_t = \frac{1}{2} 
\left[ \dot\varphi^2 + \frac{b^2_0}{b^2} \left(\nabla\varphi\right)^2
+2 V(\varphi)\right].
\end{equation}
Hence, the 4D expectation value of the Einstein equation
$\left<H^2\right> = {8\pi G \over 3} \left<\rho_t\right>$
on the 4D FRW metric (\ref{frw}), will be
\begin{equation}
\left<H^2\right> = \frac{4\pi G}{3} \left< \dot\varphi^2 +
\frac{b^2_0}{b^2} \left(\nabla\varphi\right)^2 + 2 V(\varphi)\right>,
\end{equation}
where $G$ is the gravitational constant and $\left<H^2\right>
\equiv h^2=\dot b^2/b^2$.
Now we can make a semiclassical treatment\cite{NPB}
for the effective 4D quantum field
$\varphi(\vec R,t)$, such that $<\varphi> = \phi_c(t)$:
\begin{equation}\label{semi}
\varphi(\vec{R},t) = \phi_c(t) + \phi(\vec{R},t).
\end{equation}
For consistence we take $<\phi>=0$ and $<\dot\phi>=0$.
With this approach the classical dynamics on the background 4D FRW
metric (\ref{frw}) is well described by the equations
\begin{eqnarray}
&& \ddot\phi_c + 3 \frac{\dot b}{b} \dot\phi_c + V'(\phi_c)=0, \label{ua} \\
&& H^2_c = \frac{8\pi G}{3} \left(\frac{\dot\phi^2_c}{2} + V(\phi_c)
\right),\label{hc}
\end{eqnarray}
where $H^2_c = \dot a^2/a^2$ and the prime denotes de derivative with
respect to the field.
In other words the scale factor $a$ only takes into account
the expansion due to the classical Hubble parameter, but
the effective scale factor $b$ takes into account both, clasical
and quantum contributions in the energy density: ${\dot b^2 \over b^2} =
{8\pi G \over 3} \left<\rho_t\right>$.
Since $\dot\phi_c=-{H'_c \over 4\pi G}$, from eq. (\ref{hc}) we obtain
the classical scalar potential $V(\phi_c)$ as a function of the
classical Hubble parameter $H_c$
\begin{displaymath}
V(\phi_c) = \frac{3 M^2_p }{8\pi} \left[ H^2_c - \frac{M^2_p}{12\pi}
\left(H'_c\right)^2 \right],
\end{displaymath}
where $M_p=G^{-1/2}$ is the Planckian mass.
The quantum dynamics is described by
\begin{eqnarray}
\left< H^2 \right> & = &  H^2_c + \frac{8\pi G}{3} 
\left< \frac{\dot\phi^2}{2} + \frac{b^2_0}{2 b^2} (\nabla\phi)^2 +
\sum_{n=1} \frac{1}{n!} V^{(n)}(\phi_c) \phi^n \right>.\label{fried} \\
\ddot\phi & + &  3 \frac{\dot b}{b}\phi - \frac{b^2_0}{b^2} \nabla^2\phi +
\sum_{n=1} \frac{1}{n!} V^{(n+1)}(\phi_c) \phi^n =0, \label{apr} 
\end{eqnarray}
In what follows we shall make the following identification:
\begin{equation}
\Lambda(t) = 8\pi G 
\left< \frac{\dot\phi^2}{2} + \frac{b^2_0}{2 b^2} (\nabla\phi)^2 +
\sum_{n=1} \frac{1}{n!} V^{(n)}(\phi_c) \phi^n \right>,
\end{equation}
such that
\begin{equation} \label{lam}
\frac{\dot b^2}{b^2} = \frac{\dot a^2}{a^2} + \frac{\Lambda}{3}.
\end{equation}
On cosmological scales, the fluctuations $\phi$ are small, so that
it is sufficient to make a linear approximation ($n=1$) for
the fluctuations. Thus, the second term in (\ref{lam})
is negligible on such that scales. However, the second
term in (\ref{lam}) could be important in the ultraviolet
spectrum and more exactly at Planckian scales. At these
scales the modes for $\phi$ should be coherent and the matter
inside these regions can be considered as dark.
Hence, the significative contribution for the
function $\Lambda(t)$ is given by
\begin{equation}\label{lam1}
\Lambda(t) \simeq 8\pi G 
\left.\left< \frac{\dot\phi^2}{2} + \frac{b^2_0}{2 b^2} (\nabla\phi)^2 +
\sum_{n=1} \frac{1}{n!} V^{(n)}(\phi_c) \phi^n \right>\right|_{Planck}.
\end{equation}
In this sense, we could make the identification for $\Lambda $ as
a cosmological parameter which only takes into account
the ``coherent quantum modes'' (or dark matter) contribution for the 
expectation value of energy density: $\left<\rho_{\Lambda}\right> =
\Lambda /(8\pi G)$. 

Once done the linear approximation ($n=1$) for the semiclassical treatment
(\ref{semi}), we can make the identification of the squared mass for the
inflaton field $m^2 = V''(\phi_c)$. Hence, after make a linear expansion
for $V'(\varphi)$ in eq. (\ref{a1}), we obtain
\begin{eqnarray}
V'(\phi_c) & \equiv &
\left(4\frac{h^3}{\dot h} -
3\frac{\dot h}{h} - 3 \frac{h^5}{\dot h^2}\right)
\dot\phi_c,\label{ua1} \\
m^2 \phi & \equiv  &
\left(4 \frac{h^3}{\dot h} - 3 \frac{\dot h}{h} - 
3\frac{h^5}{\dot h^2} \right) \frac{\partial \phi}{\partial t} 
+ 
\left(\frac{b^2_0}{b^2} - h^2 R^2 \right) \nabla^2_R\phi.
\label{apr1}
\end{eqnarray}
Taking into account the expressions (\ref{ua}) with (\ref{ua1}) and
(\ref{apr}) with (\ref{apr1}), we obtain the dynamics for $\phi_c$ and
$\phi$.
Hence, the equations $\ddot\phi_c+ 3h \dot\phi + V'(\phi_c)=0$
and $\ddot\phi +   3h\dot\phi - (b/b_0)^2\nabla^2_R \phi
+ V''(\phi_c)\phi=0$ now take the form\cite{MB}
\begin{eqnarray}
&& \ddot\phi_c + \left[3h+ f(t)\right]\dot\phi_c =0, \\
&&  \ddot\phi + \left[3h(t) + f(t)\right]\dot\phi -
h^2R^2\nabla^2_R\phi =0, \label{qf}
\end{eqnarray}
where
\begin{equation}\label{f}
f(t)= \left(4\frac{h^3}{\dot h} -
3\frac{\dot h}{h} - 3 \frac{h^5}{\dot h^2}\right).
\end{equation}

\subsection{Examples}

To ilustrate the formalism we shall consider two examples. The
first one is an application of
this formalism to construct a simple inflationary
model, and the second one is a proposal cosmological model in which we
consider the cosmological constant, including the inflationary era.

\subsubsection{{\bf Inflation with $\Lambda =0$}}

On cosmological scales and during inflation, the quantum fluctuations
are small, so that the linear aproximation in eq. (\ref{apr})
is sufficient to make a realistic description for
the evolution of $\phi$. Furthermore, the cosmological parameter
$\Lambda (t)$ is negligible during inflation when $\phi$ is considered 3D
spatially
homogeneous. However, such term could be important in other
times of the evolution of the universe. Taking this into account
in eq. (\ref{lam}) the effective scale factor $b(t)$ is equals to
the classical scale factor $a(t)$, and the same is for the effective
hubble parameter $h(t)$ and the classical hubble parameter $H_{c}(t)$.
During the inflationary epoch, the slow-roll condition
$\gamma(t) = -\dot H_c/H^2_c \ll 1$ holds\cite{MB1}.
Since $u(t)=1/\gamma(t)$,
we obtain that $u\gg 1$. This assures that all the velocities
in $U^A$ in (\ref{tri})
and $\hat U^A$ in (\ref{10}) to be real, and
imposes the condition $r \gg 1$\cite{MB}.
Furthermore the equation of state can be written in terms
of the function $u(t)$
\begin{displaymath}
\left<P\right> = - \left[1-\frac{2}{3 u(t)}\right] \left<\rho_t\right>,
\end{displaymath}
which, since $u \gg 1$ during inflation, complies with the required
condition for this stage: $\left<P\right> \simeq - \left<\rho_t\right>$.
Moreover, speaking in terms of the effective 4D FRW metric (\ref{frw}),
the geodesic
evolution of the fifth coordinate give us the Hubble horizon $\psi(t)=
1/H(t)$ and the resulting fifth (constant) coordinate 
$L=\psi_0$ is given by the Hubble horizon at the end of inflation:
$L=1/H_c(t_0)$.

We can define the {\it redefined quantum fluctuations} $\chi(\vec R, t)=
e^{1/2 \int [3H_c(t)+f(t)]dt} \phi$, so that the equation of motion
for $\chi$ yields
\begin{equation}
\ddot\chi - \left[H^2_c R^2 \nabla^2_R + \frac{1}{4}
\left(3H_c + f(t)\right)^2 + \frac{1}{2} \left(3\dot H_c + \dot f(t)\right)
\right]\chi =0,
\end{equation}
so that the modes $\chi_k(t)$ of the field $\chi$ complies
the differential equation
\begin{equation}\label{eq1}
\ddot\chi_k + H^2_c R^2 \left(k^2 - k^2_0(t)\right) \chi_k=0,
\end{equation}
with
\begin{equation}\label{k0}
k^2_0(t) = \frac{1}{R^2H^2_c} \left[\frac{1}{4} \left(3H_c + f(t)\right)^2
+ \frac{1}{2} \left(3\dot H_c + \dot f(t) \right)\right],
\end{equation}
where $f(t)$ is a function of the classical Hubble parameter [see eq.
(\ref{f})]. Hence, all the dynamics of the quatum fluctuations being
described only by the classical Hubble parameter $H_c=\dot a/a$.

Now we are going to study an example where
$\psi(N)=-1/(\alpha N)$, so that $H_c(N)=-\alpha N$. This implies that
the classical Hubble parameter (written as a function of time)
is given by $H_c(t)=H_0 e^{\alpha \Delta t}$. At the end of inflation
$\alpha \Delta t \ll 1$, so that $H_c(t) \simeq H_0 (1+\alpha \Delta t)$
and $3H_c(t) + f(t) \simeq 3H_0 (1+\alpha \Delta t)
+ 3\alpha -(4 H^2_0/\alpha) (1+2\alpha \Delta t)
-(3 H^3_0/\alpha^2) (1+3\alpha \Delta t)$,
where $\Delta t = t_0 -t$ and $t_0$ is the time for which inflation ends.
At the end of inflation it is sufficient to make a $\Delta t$-first order
expansion for $k^2_0$, so that it can be approximated to
\begin{equation}
k^2_0(t) = \frac{1}{r^2} \left(A_1-A_2 t\right).
\end{equation}
With this approximation, the general solution
for the modes $\chi_k(t)$ is
\begin{equation}
\chi_k(t) = C_1 {\rm Ai}\left[x(t)\right] + C_2 {\rm Bi}\left[x(t)\right],
\end{equation}
where
${\rm Ai}\left[x(t)\right]$ and ${\rm Bi}\left[x(t)\right]$
are the Airy functions with argument $x(t)$. Furthermore,
($C_1$,$C_2$) are some constants and
\begin{eqnarray}
A_1& =& \frac{1}{4} \left(3 H_0 - 3 \frac{H^3_0}{\alpha^2} + \alpha -
3\frac{H^2_0}{\alpha}\right)^2 + \frac{1}{2} \left(8 H^2_0
- 9\frac{H^3_0}{\alpha} -\alpha H_0\right) \nonumber \\
& - & \frac{1}{2}
\left(3 H_0 - 3 \frac{H^3_0}{\alpha^2} + 3\alpha -
8\frac{H^2_0}{\alpha}\right) \left(8H^2_0 + 9 \frac{H^3_0}{\alpha} -
3H_0 \alpha\right) t_0, \\
A_2  & =&  \frac{1}{2} \left(3 H_0 - 3 \frac{H^3_0}{\alpha^2} + 3\alpha
-8\frac{H^2_0}{\alpha}\right)\left(3H_0 \alpha -8 H^2_0 -
9 \frac{H^3_0}{\alpha}\right), \\
x(t) & = & \frac{\left[(A_1-k^2)-A_2 t \right]}{A_2}
\left(\frac{A_2}{r^2}\right)^{1/3}.
\end{eqnarray}
Note that in this example $H_0$ denotes the value of the Hubble parameter
at the end of inflation.
On cosmological scales (i.e., for $k^2 \ll A_1 - A_2 t$), the
solution for $\chi_k$ is unstable.
However in the UV sector (i.e., for $k^2 \gg A_1-A_2 t$),
the modes oscillate. This behavior is well described by the
function ${\rm Bi}[x(t)]$, so that we shall take
$C_1=0$. Hence, at the end of inflation the modes $\chi_k$ will
be
\begin{equation}\label{ch}
\chi_k(t) = C_2 \  {\rm Bi}[x(t)].
\end{equation}
Since the modes of the quantum fluctuations $\phi$
are $\phi_k=e^{-1/2\int [3H_c+f(t)]dt} \chi_k $, 
the squared fluctuations
are given by
\begin{equation}\label{sf}
\left<\phi^2\right> \simeq \frac{1}{2\pi^2}
e^{-\left[3(H_0+\alpha)-4\frac{H^2_0}{\alpha}
-3\frac{H^3_0}{\alpha^2}\right]t}
{\Large\int} dk \  k^2 \left|\chi^2_k\right|,
\end{equation}
where the modes $\chi_k$ are given by eq. (\ref{ch}).
Furthermore the density fluctuations at the end of inflation
can be estimated by the expression
\begin{equation}
\frac{\delta\rho_t}{\rho_t} \sim  \frac{H^2_0}{\dot\phi_c}
\sim 2 \pi^{1/2} \frac{H^{3/2}_0 }{M_p \alpha^{1/2}},
\end{equation}
which are of the order of $10^{-5}$ for $H_0 \sim 10^{-5} \  M_p$
and $\alpha \sim 10^{-5} \  M_p$.
In our case, the spectral index $n_s$ being given by
$n_s -1=-{6\over u(t)}$. During inflation $u \gg 1$, so that
$|n_s-1| \ll 1$. Hence, during inflation the spectrum approaches
very well with a Harrison - Zeldovich one.

\subsubsection{{\bf A more general cosmological model with $\Lambda \neq 0$}}

As a second example we propose a cosmological model without the above
consideration about the cosmological parameter $\Lambda$, because in this
model we are considering $\Lambda$ as a constant. That
implies the effective Hubble parameter is different to the classical Hubble
parameter ($h\neq H_c$).
Taking this into account we consider a time dependent
power expansion $p(t) = 2/3 - B t^{-1} + A t^{-2}$, such
that the classical Hubble parameter is given
by $H_c(t)=p(t)/t$ and ($A$,$B$) are
constants.
The effective power $p_1(t)$
for the effective Hubble parameter $h(t)$ will be
$p_1(t) = \sqrt{(2/3 + A t^{-2} - B t^{-1})^2 + \Lambda/3 t^2}$,
because $h^2 = H^2_c + \Lambda/3$. 
This implies that the total density parameter will be
$\Omega_T =\Omega_r + \Omega_m + \Omega_{\Lambda}=1$,
for a critical energy density given
by $\rho_t = {3 \over 8\pi G} h^2$, such that
\begin{equation}
\Omega_r + \Omega_m = \frac{H^2_c}{h^2}, \qquad \Omega_{\Lambda} = \frac{
\Lambda}{3 h^2} .
\end{equation}
where $\Omega_r$, $\Omega_m$ and $\Omega_{\Lambda}$ are respectively
the contributions for radiation, matter and $\Lambda$.
In our case, because we consider $\Omega_T = 1$, this implies that
\begin{equation}
p^2_1(t) = p^2(t) + \frac{1}{3} \Lambda t^2,
\end{equation}
where $t>0$ is the cosmic time.
We define $b/b_0 = e^N$, such that $b_0\equiv b(t=t_0)$, where
$t_0$ is the time when inflation ends (i.e., the
time for which $\ddot b=0$). Thus $N$ will be grater than zero
only for times larger than $t_0$, but negative for
$t< t_0$ (i.e., during the previous inflationary phase).
This means that the parameter $N$ give us
the number of e-folds with respect to the effective scale factor at the
end of inflation: $b_0$.
Once defined the scale for $N$, we
can see the evolution for the function $u(t)$.
During inflation $\ddot b >0$, so that $u(t) >1$
and $\epsilon = 1$. In such that
epoch $q <0$ (i.e., the universe is accelerated) and
$b/b_0=e^N < 1$, because $N <0$. In such that phase: $r \gg 1$.
This means that cosmological
scales include regions very much larger than the Hubble horizon [see
the metric (\ref{frw})].

At the end of inflation $u(t)$ take values close (but
larger) to the unity. At $t=t_0$ $\ddot b=q=0$,
the function $u(t_0)=1$, so that
the global hyperbolic geometry condition
$\hat g_{AB} \hat U^A \hat U^B=1$
it's not well defined. However, the line element (\ref{frw}) is well
defined. At this
moment the universe suffers a change of phase from a accelerated 
to a decelerated expansion and $r=1$, because $b(t=t_0)=b_0$.

During the second phase (i.e., decelerated expansion) the universe
is governed by radiation and later by matter. The function
$u^2(t)$ is smaller than the unity (but $u^2 >0$), so that
$r$ take values
${1\over 2} e^{-N} ={1\over 2} b_0/b < r < 1$, for $N >0$.
This means that, during this phase, the metric (\ref{frw}) describes
the universe on scales smaller than the Hubble radius: $r/h < 1/h$.
The interesting here is that the velocities (\ref{10}) becomes
purely imaginary and the signature of the 4-D effective
metric (\ref{frw}) changes synchronically
(with respect to the signature during the
inflationary phase): $(+,-,-,-) \rightarrow
(-,+,+,+)$; that is, $\epsilon$ jumps from $1$ to $-1$ to preserve
the global geometry in (\ref{con}).
In this sense we can say that the 4-D effective
metric (\ref{frw}) is ``dynamical''.
The fig. (1) shows the evolution of the powers $p_1[x(t)]$
(dashed line) and
$p[x(t)]$ (continuous line) as a function of $x(t) = {\rm log}_{10}(t)$ for
$A=1.5 \  10^{30} \  {\rm G}^{1}$ and $B=10^{15} \  {\rm G}^{1/2}$.
Numerical calculations give us the time for which $\ddot b =q=0$
at the end of inflation: $x(t_0)\simeq
14.778$. At this moment $N(t_0)=0$, but after it becomes positive.
Note that for $x(t) < 60.22$ both curves are very similar, but for
$x(t) > x(t_*)$ (with $x(t_*) \simeq 60.22$), $p_1$
increases very rapidly but not $p$, which remains
almost constant with a value close to $p \simeq 2/3$. The difference
between both curves is due to the presence of
a nonzero ``cosmological constant'' ($\Lambda$),
which was valued as: $\Lambda= 1.5 \  10^{-121} \  {\rm G}^{-1}$.\footnote{
At the moment the consensus has
emerged about the experimental value of the cosmological constant\cite{u,u1}.
It is on the order of magnitude of the matter energy density:
$\rho_{\Lambda} \sim (2-3)\rho_{m}$.
The Wilkinson Microwave Anisotropy Probe (WMAP) data
suggest that the universe is very nearly spatially flat,
with a density parameter $\Omega_T = 1.02 \pm 0.02$\cite{spergel}.}
In other words, at $t_* \simeq 1.66 \  10^{60} \  {\rm G}^{1/2}$
the deceleration parameter becomes zero and later negative.
At this moment, the universe changes from a 
decelerated to an accelerated phase and $\epsilon$ jumps from
$-1$ to $1$ because $u(t)$ evolves from $u(t< t_*) <1$ (decelerated
expansion) to
$u(t> t_*) > 1$ (accelerated expansion). It should be when the universe
was nearly $0.4 \  10^{10}$ years old.
The present day age of the universe was considered as $x(t)
= 60.653 \  {\rm G}^{1/2}$ (i.e., $1.5 \  10^{10}$ light years).
Note that $\Omega_r+ \Omega_m$ decreases for late times
[see figure (2)], so that
its present day value should be
$(\Omega_r+ \Omega_m)[x(t=60.653 \  G^{1/2})] \simeq 0.32$.
Thus, the present day value for the vacuum density parameter
$\Omega_{\Lambda}=1-(\Omega_r+ \Omega_m)$ should be
$\Omega_{\Lambda}[x(t=60.653 \  G^{1/2})] \simeq 0.68$.
With these parameter values
we obtain the present day deceleration parameter:
$q[x(t=60.653 \  G^{1/2})] \simeq -0.747$, so that the present day
cosmological parameter should be: $\omega[x(t=60.653 \  G^{1/2})] \simeq
-0.831$. Note that all these results are in very good agreement with
observation\cite{PDG,spergel}.\\

\noindent
Evolution of $p_1[x(t)]$ (dashed line) and $p[x(t)]$
(continuous line) as a function of $x(t) = {\rm log}_{10}(t)$, for
$A=1.5 \  10^{30} \  {\rm G}^{1}$, $B=10^{15} \  {\rm G}^{1/2}$.\\
\vskip 3cm

\noindent
Evolution of $(\Omega_m + \Omega_r)[x(t)]$
as a function of $x(t) = {\rm log}_{10}(t)$, for
$A=1.5 \  10^{30} \  {\rm G}^{1}$, $B=10^{15} \  {\rm G}^{1/2}$.\\
\vskip 3cm

\section{Extra force and extra mass}

As we saw in the previous section, it is possible to
consider a cosmological model governed by a neutral scalar field that
initially suffers an inflationary expansion that has a change
of phase towards a decelerated (radiation and later matter dominated)
expansion that thereinafter evolves towards the observed present
day (quintessential) expansion. In this section we shall study
the possibility to have fifth force and fifth mass making
an extension of the formalism developed by Ponce de Leon
to cosmological models. In particular we shall extend this formalism
to the cosmological model developed in the previous section considering
two frames. This way to obtain equations of practical use,
we can introduce the
action ${\cal S}(x^A)$ as a function of the generalyzed coordinates
$x^A$. Hence, since the momentum
${\cal P}^A = - {\partial {\cal S} \over \partial x^A}$,
for a diagonal tensor metric $g^{AB}$ we obtain the Hamilton-Jacobi
equation
\begin{equation}\label{HJ}
g^{AB}  \left(\frac{\partial {\cal S}}{\partial x^A}\right)
\left(\frac{\partial {\cal S}}{\partial x^B}\right)= M^2_{(5)},
\end{equation}
where $M_{(5)}$ is the invariant 5D gravitational
mass of the object under study (in our case, the mass of the inflaton
field).
In the particular frame (\ref{tri}), with the Lagrangian
(\ref{1}) and (\ref{1'}), $M_{(5)}$ describes the 5D mass of the
scalar field $\varphi$.
In this case the tensor metric is symmetric (and diagonal), and
the Hamilton-Jacobi equation (\ref{HJ}) adopts the particular
form
\begin{equation}
g^{NN} \left(\frac{\partial {\cal S}}{\partial \varphi_{,N}}\right)^2 +
g^{\psi\psi} \left(\frac{\partial {\cal S}}{\partial
\varphi_{,\psi}}\right)^2 =
M^2_{(5)}.
\end{equation}
On the other hand, in general, the line element (\ref{6}) can be
written as:
\begin{equation}
dS^2 = ds^2 + dS^2_{(4)},
\end{equation}
where $ds^2$ describes the 4D line element and $dS^2_{(4)}$ only
the line element related with the fifth coordinate.
We shall define the extra force
\begin{equation}
F^{ext} = \frac{d{\cal P}^{x^4}}{ds},
\end{equation}
as the force on the sub manifold
$ds^2$ due to the motion of the fifth coordinate.
In general, the momentum ${\cal P}^{x^4}$ is defined as
\begin{equation}
{\cal P}^{x^4} = \frac{\partial ^{(5)}L}{\partial \varphi_{,x^{4}}}.
\end{equation}
In the frame (\ref{tri}) ${\cal P}^{x^4} \equiv {\cal P}^{\psi}$, and
is given by
${\cal P}^{\psi} = - {\psi^4 e^{3N}\over \psi^2_0} \left(g^{\psi\psi}\right)^2
\varphi_{,\psi}$, which also can be written in terms of the potential
\begin{equation}
{\cal P}^{\psi} = -\frac{\psi^4 e^{3B}}{\psi^2_0}
g^{\psi\psi} \frac{\partial V(\varphi)}{\partial\varphi_{,\psi}}.
\end{equation}
Hence, the extra force holds
\begin{equation}
F^{ext} = \frac{\psi^3 e^{3N}}{\psi^2_0}
\left( 3 \frac{\stackrel{\star}{\psi}}{\psi} \varphi_{,\psi}
+ 3 \varphi_{,\psi}+ \stackrel{\star}{\varphi}_{,\psi}\right),
\end{equation}
where the overstar denotes the derivative with respect to $N$.

On the other hand, from the
equation $g_{AB} U^A U^B =1$, we obtain the
invariant 5D mass $M_{(5)}$
\begin{equation}
g_{AB} {\cal P}^A {\cal P}^B =M^2_{(5)},
\end{equation}
where ${\cal P}^A= M_{(5)} U^A$.
For example, in the frame (\ref{tri}) the 4D mass $m_0$ and the
5D invariant mass $M_{(5)}$ are given respectively by
\begin{equation}
M^2_{(5)}= g^{NN} \left(\frac{\partial {\cal S}}{\partial \varphi_{,N}}
\right)^2+
g^{\psi\psi} \left(\frac{\partial {\cal S}}{\partial
\varphi_{,\psi}}\right)^2 ,
\qquad m^2_0 = g^{NN} \left(\frac{\partial {\cal S}}{\partial
\varphi_{,N}}\right)^2,
\end{equation}
so that its diference
\begin{equation}
m^2_0 - M^2_{(5)} = -g^{\psi\psi}
\left(\frac{\partial {\cal S}}{\partial \varphi_{,\psi}}\right)^2,
\end{equation}
is nonzero. The interesting here is that $m^2_0 > M^2_{(5)}$. In other
words, in the frame (\ref{tri})
the motion of the fifth coordinate has an antigravitational effect
on the field $\varphi$ in the submanifold (or bulk) $ds^2$.
However, this frame is not very instructive because $N$ and $r$
are not physical coordinates. Next we shall study some
examples which could be relevant in cosmological models. The first one
is the case of the cosmological model developed previously seen from the
frame  $(t,R,L)$ defined by the speeds (\ref{10}). It is
easy to see that in this frame the 5D momentum
${\cal P}^L$ is null: ${\cal P}^L=0$.
This implies that the extra force will be
\begin{equation}
F_{ext} = 0.
\end{equation}
It also can be viewed from the point of view of the extra mass.
In this frame  $m^2_0 = M^2_{(5)}$ where
\begin{equation}
\left(\frac{\partial {\cal S}}{\partial\varphi_{,t}}\right)^2 -
e^{2\int h(t) dt}
\left(\frac{\partial {\cal S}}{\partial\varphi_{,R}}\right)^2=
M^2_{(5)}.
\end{equation}
Hence, the inertial 4D mass $m_0$ is the same than the invariant
5D mass $M_{(5)}$, so that there is not extra force on the
effective 4D frame. This can be justified from the fact that
the fifth coordinate $L$ do not varies in this frame. In other words
the 4D bulk $ds^2$ is the same that the 5D manifold $dS^2$, because
$dS_{(4)} =0$ for an observer that ``expands with the universe'' in
an inertial frame.\\

Other interesting frame it is that whose fifth coordinate is variable. This
can be described by means of the transformation
$t = \int \psi(N) dN$, $R=r\psi $ and $\xi=\psi(N) e^{\int
\stackrel{\star}{H(N)}/H(N) dN}$, so that the 5D velocities
are
\begin{eqnarray}
&& U^t = \frac{2u(t)}{\sqrt{u^2(t) -1}}, \label{f1} \\
&& U^R = -\frac{2r}{\sqrt{u^2(t) -1}}, \label{f2} \\
&& U^{\xi} = \frac{u(t)}{\sqrt{u^2(t) -1}} \left(\frac{\dot H}{hH}
-\frac{\dot h}{h^2}\right) \frac{H}{H_0}. \label{f3}
\end{eqnarray}

In this frame
the 5D line element is given by\cite{MB2}
\begin{equation}\label{met}
dS^2 = \epsilon\left[
dt^2 - e^{2 \int h(t) dt} dR^2 - \left(\frac{H_0}{H}\right)^2 d\xi^2\right],
\end{equation}
where the 4D line element (or ``bulk'') $ds^2$ is given by
the first two terms in (\ref{met})
\begin{equation}\label{bulk}
ds^2 = \epsilon\left( dt^2 - e^{2 \int h(t) dt} dR^2\right),
\end{equation}
and $h^2(t) = H^2(t) + {C\over 3}$ for a given constant $C$.
Hence, the extra force on the 4D bulk
will be $F^{ext} = {d{\cal P}^{\xi} \over ds}$.
Note that extra force becomes from the motion of the fifth coordinate
in the effective 4D bulk. In other words, an observer in the 4D bulk
(\ref{bulk}), will move under the influence of an extra force
that, in the example here studied, takes the form
\begin{equation}\label{for}
F^{ext} = 
\left(\left|1- \frac{r^2 \dot h^2}{h^4} e^{2\int h dt} \right|\right)^{-1/2}
\frac{d{\cal P}^{\xi}}{dt},
\end{equation}
which is invariant under changes of signature (i.e., $\epsilon=1 \rightarrow
\epsilon=-1$).
The 5D Lagrangian in this frame takes the form
\begin{equation}\label{lag}
^{(5)}L(\varphi,\varphi_{,A})
= - \left(\frac{b}{b_0}\right)^3
\frac{H_0}{H}
\left(\frac{1}{2} g^{\alpha\beta} \varphi_{,\alpha}\varphi_{,\beta}
+ V(\varphi) \right),
\end{equation}
so that the momentum ${\cal P}^{\xi}$ is
\begin{equation}
{\cal P}^{\xi} = - \left(\frac{b}{b_0}\right)^3
\frac{H_0}{H} g^{\xi\xi}
\frac{\partial V(\varphi)}{\partial\varphi_{,\xi}}.
\end{equation}
In this representation the potential $V(\varphi)$ assumes the form
\begin{equation}\label{pote}
V(\varphi) = \frac{1}{2} \left[\left(R h\right)^2
-\left(\frac{b_0}{b}\right)^2
\right] \left(\nabla_{R} \varphi\right)^2 -
\frac{1}{2} \left(\frac{H}{H_0}\right)^2 \varphi^2_{,\xi} -
\frac{H}{H_0} \left(R h\right)\varphi_{,\xi} \nabla_{R} \varphi,
\end{equation}
so that the momentum $P^{\xi}$ is
\begin{equation}\label{p}
P^{\xi} =
\left(\frac{b}{b_0}\right)^3 \left[\left(\frac{H}{H_0} \right)
\varphi_{,\xi} + \left(R h\right) \nabla_{R}\varphi\right].
\end{equation}
Note that the effective kinetic component in the 5D Lagrangian
(\ref{lag}) is 4D, but the potential (\ref{pote}) is evaluated
in the 5D frame (\ref{f1}),(\ref{f2}),(\ref{f3}).
From eqs. (\ref{for}) and (\ref{p}), we obtain the extra force for
this frame
\begin{eqnarray}
F^{ext}& = & \left(\frac{b}{b_0}\right)^3 \left[\left|1- \left(\frac{
R \dot h}{h}\right)^2 \left(\frac{b}{b_0}\right)^2\right| \right]^{-1/2}
\left[\left(3\frac{\dot b}{b} \frac{H}{H_0} + \frac{\dot H}{H} \right)
\varphi_{,\xi} + \left( 3 \frac{\dot b}{b} \left(R h\right) +
\left(\dot R h + R \dot h\right)\right) \nabla_{R}\varphi\right. \nonumber \\
&+& \left.\frac{H}{H_0} \frac{d}{dt}\left(\varphi_{,\xi}\right) +
\left(R h\right) \frac{d}{dt} \left(\nabla_r\varphi\right)\right], \label{fo}
\end{eqnarray}
where $\left({b\over b_0}\right)^2 = e^{2\int h dt}$.
The extra force is originated in the
last two terms of the 5D potential (\ref{pote}), which depends on the fifth
coordinate $\xi$.

On the other hand the 4D squared mass of the inflation field $\varphi$
on the 4D bulk (\ref{bulk}), is given by
\begin{equation}
m^2_0 = \left(\frac{\partial {\cal S}}{\partial\varphi_{,t}}\right)^2 -
e^{-2\int h dt} \left(\frac{\partial {\cal S}}{\partial\varphi_{,R}}\right)^2,
\end{equation}
so that one obtains
\begin{equation}\label{ma}
m^2_0 - M^2_{(5)} = \left(\frac{H}{H_0}\right)^2
\left(\frac{\partial {\cal S}}{\partial\varphi_{,\xi}}\right)^2,
\end{equation}
which gives $m^2_0 \ge M^2_{(5)}$ because the right hand of the
equation (\ref{ma}) is positive (for $C >0$). This is an important
result which shows that the motion of the fifth coordinate has
an antigravitational effect on an observer in a 4D bulk in
which the inflaton field has a 4D mass $m_0$.
This fact should be responsible for the extra force (\ref{fo})
because the observer ``is placed'' in a non inertial
frame (or 4D bulk).
In this framework the motion of the fifth coordinate
is viewed on the bulk as an extra force.
Note that it becomes zero as $C \rightarrow 0$,
because in this limit $U^{\xi} \rightarrow
0$ and $V(\varphi) \rightarrow {1 \over 2} \left[
\left(R h\right)^2
-\left({b_0\over b}\right)^2 \right] \left(\nabla_{R} \varphi
\right)^2$.
On the other hand, $U^{\xi} \rightarrow 0$ as $t\rightarrow \infty$,
because $\dot H <0$ (and $\dot h <0$)
along all the history of the universe, such that
$\left({H\over H_0}\right)_{t\rightarrow \infty} \rightarrow 0$.
Hence, for very late times the external force (\ref{fo})
on the bulk becomes
negligible. However, this force should be very important in the early
universe when $H/H_0 \gg 1$ (note that $H_0$ is the value of the
Hubble parameter at the end of inflation).

To ilustrate the formalism we can consider the case where
$h(t) = t^{-1} \  p_1(t)$ and $H(t) = t^{-1} \  p(t)$, where
\begin{eqnarray}
&& p_1(t) = \sqrt{(2/3 + A t^{-2} - B t^{-1})^2 + \frac{C}{3} t^2}, \\
&& p(t) =  \sqrt{2/3 - B t^{-1} + A t^{-2}}
\end{eqnarray}
Here
$A=1.5 \  10^{30} \  {\rm G}^{1}$, $B=10^{15} \  {\rm G}^{1/2}$
and we take the special case where the constant
$C$ is the cosmological constant $\Lambda$: $\Lambda
= 1.5 \  10^{-121} \  {\rm G}^{-1}$.
Furthermore,
$G=M^{-2}_p$ is the gravitational constant and $M_p=1.2 \times
10^{19} \  {\rm GeV}$ is the
Planckian mass.
Numerical calculations give us the time for which $\ddot b =q=0$
at the end of inflation: $x(t_0)\simeq
14.778$ [we take
$x(t) = {\rm log}_{10}(t)$].
At this moment $N(t_0)=0$, but after it becomes positive.
Furthermore, for
$x(t) > x(t_*)$ [with $x(t_*) \simeq 60.22$], $p_1$
increases from the value $p_1 \simeq 2/3$
and the 4D bulk universe is accelerated.\\

\section{Conclusions}

In this work we have studied a model for the evolution of the
universe which is globally described by a single scalar field from
5D apparent vacuum. Such vaccum is described by a flat 5D
metric with coordinates
($N$,$r$,$\psi$) and a Lagrangian for a free and minimally coupled to
gravity scalar field.
The interesting is that the scalar potential $V(\varphi)$
appears in the 3D comoving frame characterized by $U^r=0$
[see eq. (\ref{v1})].
A further transformation to physical coordinates
$t=\int \psi(N) dN$, $R=r\psi$ and $L=\psi e^{-\int dN/u(N)}$ give
us the possibility to describe the system in an effective 4D
(but 3D spatially flat) FRW metric.
Such that metric is viewed as a particular frame (characterized with $U^L=0$),
where the potential
$V(\varphi)$ is represented as the differential operator
(\ref{au}). In other words, the potential,
which assume different representations in different
frames, has a geometrical origin.
Moreover, the mass of the inflaton field
appears in the frame $U^L=0$ as a differential operator applied
to the quantum fluctuations $\phi(\vec R,t)$. Hence, for the
semiclassical treatment here developed,
$m^2\phi(\vec R,t)$
is a local operator with nonzero expectation value.
At this point we must to exalt this result, because
a particular frame in physics is intrinsically related
to an observer (or experimental result).

This 5D formalism could be extended to other particular frames
or quantum fields. Moreover, the evolution of the universe could
be examined taking into account also electromagnetism by introducing
off-diagonal terms in the metric (\ref{6}), which should be relevant to
study 3D spatial anisotropies in the universe on astrophysical
scales. However, all these issues go beyond the scope of this work.
Another important aspect that we have studied in this work is
the possible origin of extra force and extra mass
from a noncompact Kaluza-Klein formalism
by using the Hamilton-Jacobi formalism in the
framework of cosmological models.
We have examined the inertial
4D mass $m_0$ of the inflaton field
on a 4D FRW bulk in two examples.
In the first one there is not motion of the fifth coordinate with
respect to the 4D FRW bulk, so that the inertial mass $m_0$ is the same
than the 5D gravitational mass $M_{(5)}$ of the inflaton field.
As consequence of this fact there is not extra force on the 4D bulk $ds^2$
because $dS^2 = ds^2$.
However, in the second example
antigravitational effects on a non inertial 4D bulk
should be a consequence of the motion of the fifth coordinate with
respect to this bulk, because $dS^2 \neq ds^2$ so that $m^2_0 > M^2_{(5)}$.
This disagreement between the 4D inertial and 5D
gravitational masses is viewed
on the 4D bulk as an extra force. The important here is that $m_0$ has a
geometrical origin and depends on the frame of the observer. However,
$M_{(5)}$ is a 5D invariant gravitational mass and do not depends
on the frame of the observer.
This is the same situation as in the Randall-Sundrum brane-world
scenario\cite{7,8}
and other noncompact Kaluza-Klein theories, where the motion of test
particles is higher-dimensional in nature.
In other words, all test particles travel on five-dimensional geodesics
but observers, who are bounded to spacetime, have access only to
the 4D part of the trajectory.
Finally, in the cosmological model here studied, we find that both,
the discrepance between $m_0$ and $M_{(5)}$ and extra force, are bigger
in the early universe [i.e., during inflation
($x(t) < 14.778$)], but become negligible for large (present day)
times.\\

\end{document}